\documentclass[12pt, preprint]{iopart}
\usepackage[utf8]{inputenc}
\usepackage[T1]{fontenc}
\usepackage{subfigure}
\usepackage{graphicx}
\usepackage{siunitx}
\usepackage{booktabs}
\usepackage{fancyhdr}
\usepackage{xcolor}

\usepackage[colorlinks, citecolor = blue, urlcolor = blue]{hyperref}
\usepackage[backend=biber, style=chem-angew, articletitle=true]{biblatex}
\newcommand\shorttitle{Absolute frequency measurement of a Yb optical clock}

\begin{document}
\pagestyle{fancy}
\fancyhead{}
\fancyhead[L]{\shorttitle}
\title{Absolute frequency measurement of a Yb optical clock at the limit of the Cs fountain}

\author{ Irene Goti$^{1,2}$, Stefano Condio$^{1,2}$, Cecilia~Clivati$^1$, Matias~Risaro$^1$,  Michele Gozzelino$^1$, Giovanni~A. Costanzo$^{1,2}$, Filippo~Levi$^1$, Davide Calonico$^1$ and Marco Pizzocaro$^{1}$}

\address{$^1$ Istituto Nazionale di Ricerca Metrologica (INRiM), Strada delle Cacce 91, 10135 Torino, Italy}
\address{$^2$ Dipartimento di Elettronica e Telecomunicazioni, Politecnico di Torino, Corso duca degli Abruzzi 24, 10129 Torino, Italy}

\ead{m.pizzocaro@inrim.it}

\begin{abstract}
We present the new absolute frequency measurement of ytterbium ($^{171}$Yb) obtained at INRiM with the optical lattice clock IT-Yb1 against the cryogenic caesium ($^{133}$Cs) fountain IT-CsF2, evaluated through a measurement campaign lasted 14 months. Measurements are performed by either using a hydrogen maser as a transfer oscillator or by synthesizing a low-noise microwave for Cs interrogation using an optical frequency comb. The frequency of the $^{171}$Yb unperturbed clock transition ${^1}$S$_0\rightarrow {^3}$P$_0$ results to be \SI{518295836590863.44(14)}{Hz}, with a total fractional uncertainty of \num{2.7e-16} that is limited by the uncertainty of IT-CsF2. Our measurement is in agreement with the Yb frequency recommended by the Consultative Committee for Time and Frequency (CCTF). This result confirms the reliability of Yb as a secondary representation of the second and is relevant to the process of redefining the second in the International System of Units (SI) on an optical transition. 
\end{abstract}

\noindent{\it Keywords\/}
: frequency metrology, optical lattice clock, Cs fountain, SI second\\
\\
\submitto{\MET}

\section{Introduction}
The definition of the SI second is currently realized through a Cs frequency transition in the microwave regime~\cite{Takamizawa2022, Beattie2020, Weyers2018, Jallageas2018, Guena2017, Blinov2017, Fang2015, Levi2014, Guena2014, Guena2012, Li2011}. Nevertheless, in recent years it has been widely demonstrated that optical clocks represent a better frequency standard~\cite{Beloy2021, Brewer2019, McGrew2018, Ushijima2015}, both in terms of fractional frequency instability and estimated systematic uncertainty. Specific optical transitions are currently indicated by the Consultative Committee for Time and Frequency (CCTF) as secondary representations of the SI second, and recent progress in the field of optical clocks paves the way to a redefinition of the SI second~\cite{Gill2016, Lodewyck2019, Riehle2018}. 
In 2022, the CCTF published a roadmap stating relevant milestones to be achieved to proceed to a redefinition of the SI second, among which emerge multiple and independent measurements of optical clocks relative to independent Cs primary clocks.
One of the secondary representations of the second recommended by the CCTF is the transition $^1$S$_0 \rightarrow ^3$P$_0$ of $^{171}$Yb. Absolute frequency measurements of this clock transition have been realized by several metrological institutes worldwide~\cite{McGrew2018, Lemke2009, Kohno2009,  Yasuda2012, Park2013, Kim2017, Kobayashi2020, Luo2020, Kim2021, Clivati2022a}. At INRiM we measured the absolute frequency of the $^{171}$Yb clock (IT-Yb1) relative to our local cryogenic Cs fountain (IT-CsF2)~\cite{Pizzocaro2017} and via International Atomic Time (TAI)~\cite{Pizzocaro2020}. Moreover, we measured the frequency ratio between $^{171}$Yb and $^{87}$Sr clock with a transportable optical Sr clock developed at PTB~\cite{Grotti2018}, and via VLBI (Very Long Baseline Interferometry) in collaboration with NICT INAF and BIPM~\cite{Pizzocaro2021}. Recently we used the new fiber link between INRiM and SYRTE to measure the Yb clock frequency against the french Cs-Rb fountain~\cite{Clivati2022a}

In this article, we report the improvements of our optical lattice clock IT-Yb1 and the results obtained from a new measurement of the Yb absolute frequency against IT-CsF2 over a period of 14 months. The frequency measurement is performed with two different techniques: with the first method, the hydrogen maser is used as a transfer oscillator between the optical clock and the fountain, while with the second method the microwave interrogating the fountain is directly obtained by photonic synthesis on a frequency comb, using the Yb clock laser as a local oscillator.

This paper is organized as follows: in section~\ref{Expsetup} we describe the main components of our experimental setup, i.e. the Yb clock, the Cs fountain and the two measurement techniques. In section~\ref{Results} we report the results obtained in terms of Yb absolute frequency. Finally, section~\ref{Conclusions} presents the conclusions of our work.

\section{Experimental setup}
\label{Expsetup}
\subsection{IT-Yb1 and IT-CsF2}
IT-Yb1 is the optical lattice clock based on $^{171}$Yb atoms developed at INRiM. The uncertainty budget of IT-Yb1 during this measurement campaign is reported in Table~\ref{YbUnc} and the total relative uncertainty is \num{1.9e-17}. The typical instability of IT-Yb1 is $\num{2e-15}/\sqrt{\tau/\si{s}}$. IT-Yb1 is among the optical clocks that regularly submit data to the International Bureau of Weights and Measures (BIPM) for the calibration of the International Atomic Time (TAI)~\cite{Pizzocaro2020, Kim2021, Nemitz2021, Kobayashi2020, Vallet2018}.

\begin{table}[h!]
\caption{\label{BudgetYb}Uncertainty budget Yb optical lattice clock at INRiM. Other shifts are AOM switching, fibre links, and line pulling.}
\centering
\label{YbUnc}

\begin{tabular}{lcc}
\toprule
Effect 	&{Rel.\ Shift/\num{e-17}}	&{Rel.\ Unc./\num{e-17}}\\
\midrule
Density shift	& -0.5	& 0.2\\
Lattice shift	& 0.8	& 1.2\\
Zeeman shift	& -3.12	& 0.02\\
Blackbody radiation shift (room)	& -234.9	& 1.2\\
Blackbody radiation shift (oven)	& -1.3	& 0.6\\
Static Stark shift	& -1.7	& 0.2\\
Probe light shift	& 0.04	& 0.03\\
Background gas shift	& -0.5	& 0.2\\
Servo error	& 0.0	& 0.3\\
Other shifts	& 0.0	& 0.1\\
Grav. redshift	& 2599.5	& 0.3\\
\midrule
Total	& 2358.3	& 1.9\\
\bottomrule
\end{tabular}
\end{table}

IT-Yb1 and its details have been described in previous work~\cite{Pizzocaro2017, Pizzocaro2020}. An ultrastable laser at \SI{1156}{nm} is frequency duplicated to generate the \SI{578}{nm} light that is kept resonant with the $^{171}$Yb clock transition $^1$S$_0 \rightarrow ^3$P$_0$. A new ultrastable cavity directly stabilizes the \SI{1156}{nm} radiation instead of its second harmonic at \SI{578}{nm}. It is a horizontal 10-cm-long cavity, made in ultra-low-expansion glass and fused silica mirror, characterized by a flicker floor of $\sim$ \num{1e-15}. Compared to previous publications we improved the evaluation of the lattice shift and the static stark shift. In particular, we changed the optical setup from a horizontal to a vertical optical lattice with the advantage of having the strong axis of the trap contrasting the gravity. In this way, we managed to reduce the minimum trapping depth to 75 $E_\text{r}$, where $E_\text{r}\simeq h\times 2$ kHz is the recoil energy for Yb and $h$ is the Plank constant. We also decreased the depth of our working point from 200 $E_\text{r}$~\cite{Pizzocaro2017} to 100 $E_\text{r}$, with consequent advantages in terms of uncertainty budget. The working point of the lattice frequency is chosen at $\nu_{lattice} = \SI{394 798 267}{MHz}$, and it is continuously measured by an optical comb. Moreover, atoms are loaded into the lattice directly at the working depth $U$, making the atomic temperature of the trapped atoms linear as a function of the lattice depth. It is then possible to apply the approach described in~\cite{Brown2017, Beloy2020} and calculate the relative lattice shift by following the model: $\frac{\delta\nu}{\nu}=-\alpha U-\beta U^2$, where $\alpha$ and $\beta$ are two parameters that have to be measured in the specific experimental conditions. We estimate a model uncertainty of \num{1e-17}~\cite{Brown2017} resulting from deviation from linearity between the atomic temperature and the lattice depth.

We note that, since the density shift is smaller for lower lattice depth~\cite{Nemitz2019}, at the new working point of 100 $E_\text{r}$ the density shift is reduced by a factor of 10 if compared to previous work~\cite{Pizzocaro2020}. Furthermore, the atomic tunnelling between adjacent lattice sites is suppressed in a vertical lattice~\cite{Lemonde2005, McGrew2018}, making this source of uncertainty negligible.

Eight electrodes are placed on the vertical windows, four on the top window and four on the bottom, to evaluate the static stark shift. By applying voltages of up to $\pm$\SI{150}{V} independently to each electrode, it is possible to measure its contribution along the three spatial directions~\cite{Sherman2012, Lodewyck2012}. With this new setup, we detect a stray electric field of about 1-2 V/cm on the horizontal plane, which is found to change slightly from day to day. The direction of the field suggests that the oven used to generate the Yb atomic beam emits some ionized particles that accumulate on the windows. We manage to compensate for the stray electric field by using the same electrodes. The electric field in the 3 directions is measured typically 3 times per week. As each measurement is independent, we allow the resulting shift to be averaged over the campaign. The resulting shift for the period considered in this work is reported in Table~\ref{BudgetYb}.\\

IT-CsF2 is a cryogenic Cs fountain~\cite{Levi2014} and since 2013 it operates as a primary frequency standard contributing to the realization of UTC(IT) and TAI, with over 50 TAI calibration campaigns performed and published on the BIPM circular T. The uncertainty budget of IT-CsF2 is reported in Table~\ref{BudgetCs} and the total relative uncertainty is \num{2.3e-16}.

The density shift is evaluated by alternating clock measurements with 2 different atom numbers and extrapolating to zero-density. Indeed, the knowledge of the collisional-shift parameter for performing the extrapolation progressively improves with time, as its computation takes advantage of using historical and newly accumulated data, until some parameter characterizing the shift does change, requiring a new measurement run to re-calibrate the density-shift extrapolation.

In its usual configuration, the local oscillator for this clock is a BVA quartz
phase-locked to an H-maser and upconverted to \SI{9.2}{GHz} using cascaded multiplication stages and a Dielectric Resonator Oscillator (DRO)~\cite{Levi2014}. This signal is tuned to the atomic resonance using a Direct Digital Synthesizer (DDS) at about \SI{7.4}{MHz}, adjusted at each interrogation cycle. The typical short-term stability of IT-CsF2 results from a combination of the Dick effect due to the BVA quartz noise and the atomic background noise due to hot Cs atoms. It ranges between $\num{2.4e-13}/\sqrt{\tau/\si{s}}$ in high density regime and $\num{2.8e-13}/\sqrt{\tau/\si{s}}$ at low density.

\begin{table}[h!]
\caption{\label{BudgetCs}Uncertainty budget of IT-CsF2.}
\centering
\label{BudgetCs}

\begin{tabular}{lcc}
\toprule
Effect 	&{Rel.\ Shift/\num{e-16}}	&{Rel.\ Unc./\num{e-16}}\\
\midrule
Zeeman shift    & 1080.3 & 0.8\\
Blackbody radiation & -1.45 & 0.12\\
Gravitational redshift  & 260.4 & 0.1\\
Microwave leakage   & -1.2 & 1.4\\
DCP (Distributed Cavity Phase)& - & 0.2\\
Second order cavity pulling & - & 0.3\\
Background gas  & - & 0.5\\
Density shift   & -19.8 & 1.6\\
\midrule
Total   & 1318.2 & 2.3\\
\bottomrule
\end{tabular}

\end{table}

\subsection{The measurement chain}
\begin{figure}[h]
    \centering
    \includegraphics[width=\textwidth]{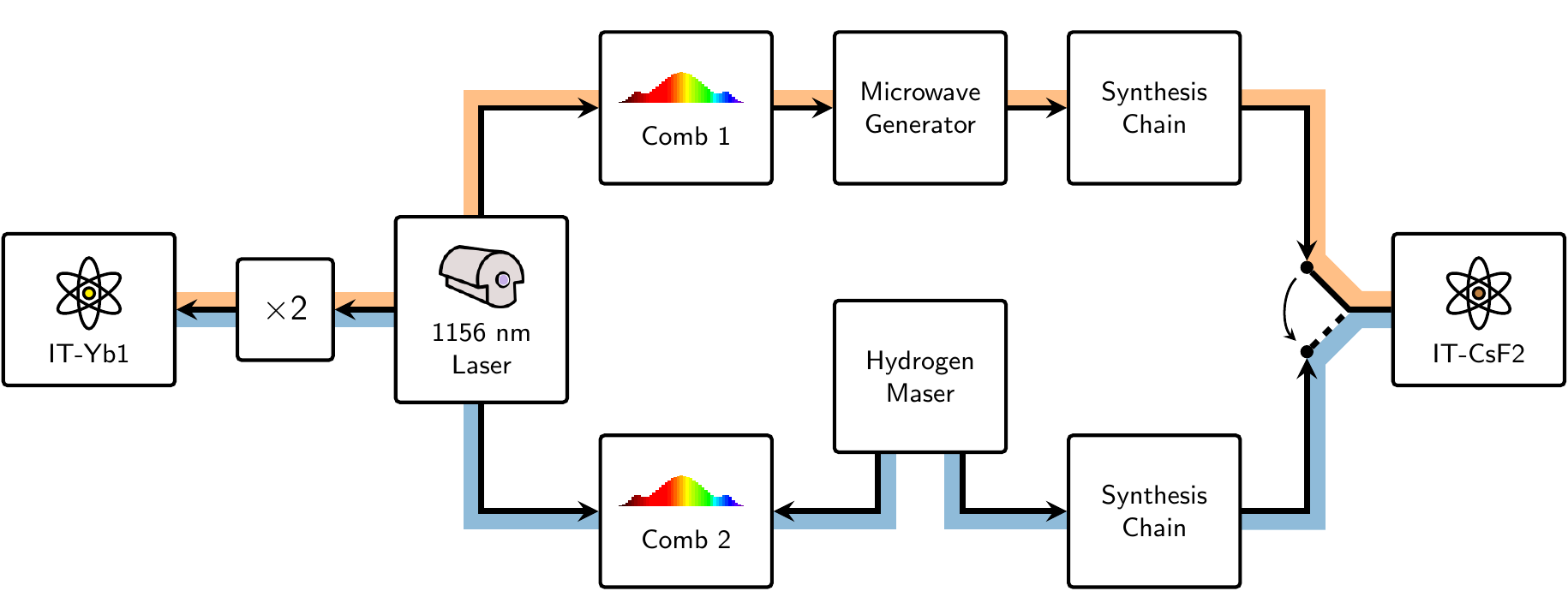}
    \caption{Experimental setup scheme of IT-Yb1 absolute frequency measurement against a IT-CsF2 clock. In blue is represented the path of the H-maser chain, which is our usual measurement chain. In orange instead is shown the microwave-to-optical chain: a new measurement technique we tested in this campaign.}
    \label{fig:scheme}
\end{figure}

Figure~\ref{fig:scheme} shows the block diagram of the two measurement chains used to compare IT-Yb1 and IT-CsF2.
The usual measurement scheme is represented in the lower branch (blue path, hereafter referenced to as "H-maser chain"). Radiation at \SI{1156}{nm}, frequency-stabilised to the ultrastable cavity, is doubled to \SI{578}{nm} for interrogating the $^{171}$Yb clock transition. Part of it is also beaten to a fibre frequency comb stabilised to the same BVA quartz/H-maser system that serves as a local oscillator for IT-CsF2. 
The IT-Yb1/IT-CsF2 ratio can thus be computed using the comb and H-maser as transfer oscillators. 
Some of the measurements are instead collected using the measurement chain shown in Fig.~\ref{fig:scheme} with the orange path (hereafter referred to as "Optical Microwave chain"). In this configuration, the frequency comb is referenced to the ultrastable \SI{1156}{nm} radiation and operates as a low-noise microwave generator to synthesize a signal resonant with the Cs clock transition. The coherent spectral transfer between optical and microwave frequencies is obtained by tightly phase-locking the comb to the optical reference (bandwidth $\approx $ \SI{300}{kHz}), following a scheme similar to the one described in~\cite{Millo2009}. The comb pulse train is collected by a fast photodiode, where the 40th harmonic of the repetition rate at \SI{10}{GHz} is selected and amplified. We note that coherent transfer of spectral properties of an optical oscillator to a microwave could alternatively be designed for optimized robustness, without recurring to a fast optical comb lock~\cite{Lipphardt2017} as we did. The noise of the generated microwave is dominated by a flicker phase process at the level of \SI{-105}{dBc/Hz} at \SI{1}{Hz}, limited by detection and signal conditioning stages. The Cs interrogation signal is synthesized from the \SI{10}{GHz} input by a two-stage down-conversion. In the first stage, we produce a \SI{9.2}{GHz} microwave by mixing to a \SI{800}{MHz} radio-frequency obtained by downscaling the input \SI{10}{GHz} signal. This is then brought to resonance with the Cs transition by mixing to a DDS at about \SI{7.4}{MHz}, adjusted at each clock cycle. The flicker phase noise contributed by this chain is \SI{-106}{dBc/Hz} at \SI{1}{Hz} Fourier frequency, corresponding to an instability of \SI{4e-16}{} at \SI{1}{s}, with a temperature sensitivity of \SI{0.07}{ps/K}. 
The ultrastable \SI{1156}{nm} laser that is used as a seed for the microwave synthesis is also calibrated by IT-Yb1 during the clock operation.

\section{Results}
\label{Results}
\subsection{Absolute frequency measurement of Yb}

As usual with atomic clocks, data is processed and reported as fractional frequencies $y = r/r_0 - 1$, where $r$ is the frequency measurement and $r_0$ is an arbitrary reference frequency.
For this analysis, the reference is chosen consistent with the recommended value of the secondary representation of the second for the Yb transition, i.e. \SI{518295836590863.63}{Hz}, which has an uncertainty of \num{1.9e-16} as established by CCTF in 2021~\cite{CCTF2021}. The use of fractional frequency linearizes equations and is convenient for computations with numbers with many digits~\cite{Lodewyck2020, Margolis2015, Gill2016}.

\begin{figure}[h]
\begin{subfigure}(a)
\includegraphics[width=9.5cm]{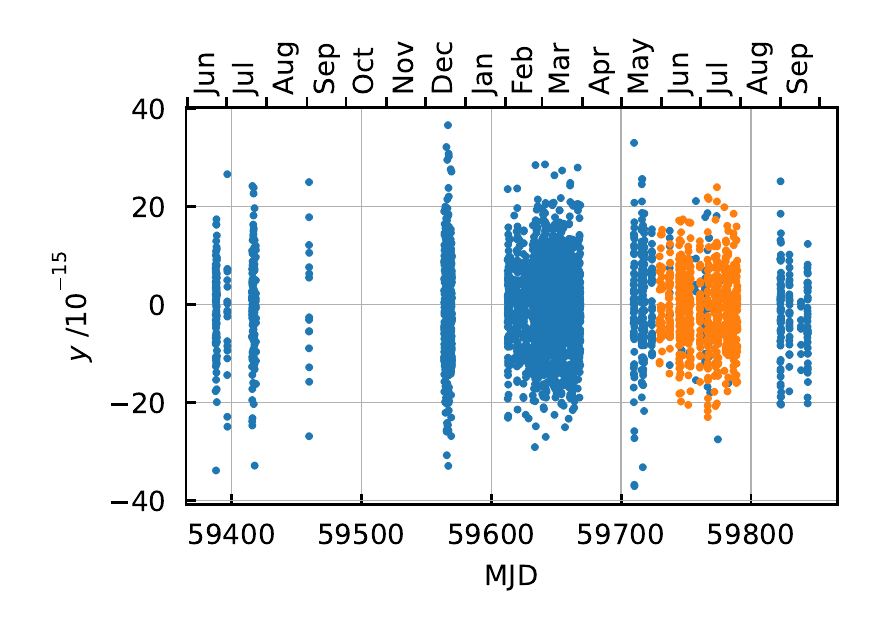}
\end{subfigure}

\begin{subfigure}(b)
\includegraphics[width=9.5cm]{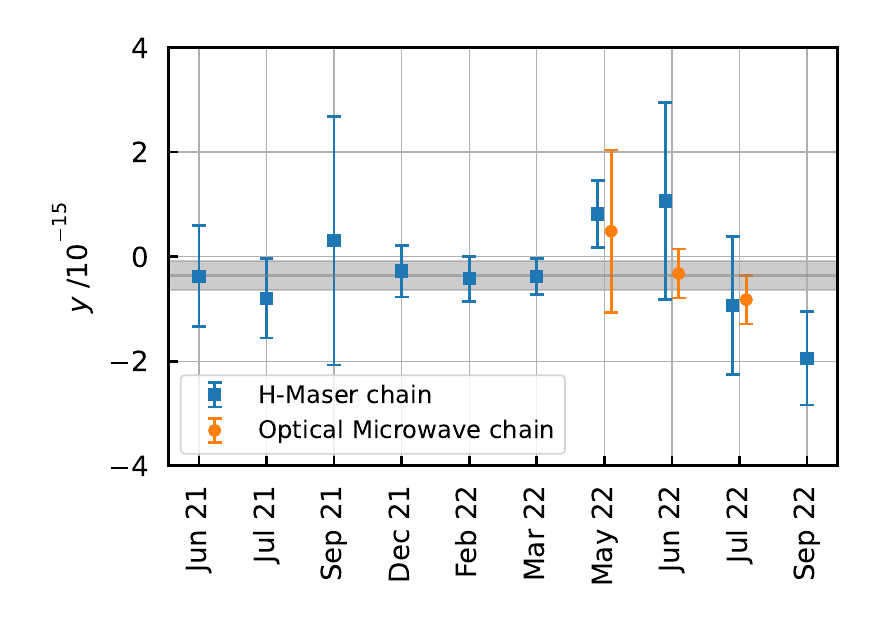}
\end{subfigure} 
\caption{a) Fractional frequency ratio of IT-Yb1 and IT-CsF2 reported as $y$ (see main text). Data were collected from June 2021 to September 2022 using the H-maser chain (blue dots) and the Optical Microwave chain (orange dots). b) Same data averaged monthly. The grey line represents the average fractional frequency ratio and the shaded area its uncertainty.}
\label{fig:data}
\end{figure}
We collected data from June 2021 to September 2022. Within this period, we have collected a total of 32 days of data with common uptime between IT-Yb1 and IT-CsF2 using the H-maser chain and a total of 6.9 days of data using the Optical Microwave chain. Data from the combs and IT-Yb1 are collected with a sampling time of \SI{1}{s}. Data from IT-CsF2, either the frequency measurement relative to the maser or the optically generated microwave, are recorded with a sampling time of \SI{864}{s}. Data from the optical clock and optical combs are averaged in bins with a duration of \SI{864}{s} and combined with the Cs fountain data. The recorded data collected in these \SI{864}{s} bins is shown in Fig.~\ref{fig:data}a, while Fig.~\ref{fig:data}b shows the same data averaged monthly. The shaded area represents the average fractional frequency ratios over the entire campaign and its uncertainty. All uncertainties include the systematic uncertainties of the clocks.
\begin{figure}[h]
    \centering
    \includegraphics[width=12cm]{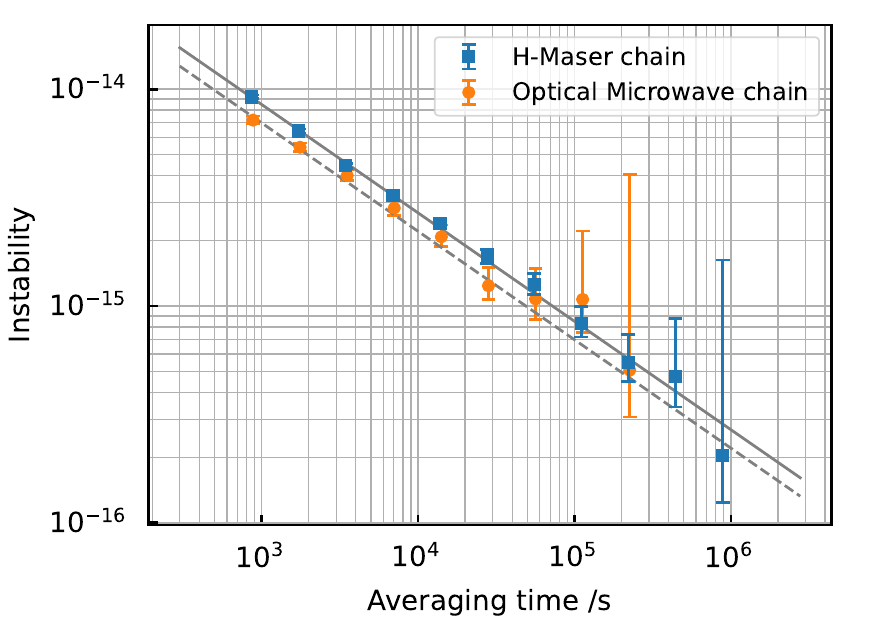}
    \caption{Overlapping Allan deviation of the comparisons over the entire campaign. Orange(blue) dots show data of a measurement carried out using Optical Microwave(H-maser) chain and the dashed(solid) line represents its instability at $\num{2.1e-13}/\sqrt{\tau/\si{s}}$ ($\num{2.7e-13}/\sqrt{\tau/\si{s}}$).}
    \label{fig:instabilities}
\end{figure}

The instabilities of the comparisons for the entire campaign are shown in Fig.~\ref{fig:instabilities} as overlapping Allan deviations. The comparison through the H-maser chain has an instability at $\num{2.7e-13}/\sqrt{\tau/\si{s}}$ over the full campaign. The comparison through the Optical Microwave chain has slightly lower instability at $\num{2.1e-13}/\sqrt{\tau/\si{s}}$. The improvement is obtained thanks to the lower phase noise of the optically-generated microwave compared to that introduced by the BVA quartz/H-maser system. We have estimated that the contribution of the Dick effect is reduced from \SI{1.6e-13}{}$/\sqrt{\tau/s}$ to a value lower than \SI{3e-15}{}$/\sqrt{\tau/s}$. The best instability observed over a short period of time (less than 1 day) is reduced from \SI{2.1e-13}{}$/\sqrt{\tau/s}$ to \SI{1.3e-13}{}$/\sqrt{\tau/s}$ in high density regime. The residual instability is dominated by the atomic background noise. The average fractional frequency difference between IT-Yb1 and IT-CsF2 is $y(\text{IT-Yb1/IT-CsF2, Optical Microwave chain}) = -3.2(3.7)\times 10^{-16}$ as measured with Optical Microwave chain and $y(\text{IT-Yb1/IT-CsF2, H-maser chain}) = -3.8(2.9)\times 10^{-16}$ measured with H-maser chain. The statistical uncertainties are respectively \num{2.8e-16} and \num{1.6e-16}, derived from the instabilities shown in Fig.~\ref{fig:instabilities}. The weighted average of the two measurements is $y(\text{IT-Yb1/IT-CsF2}) = -3.6(2.7)\times 10^{-16}$, that corresponds to an absolute frequency measurement of $f(\text{IT-Yb1}) = \SI{518295836590863.44(14)}{Hz}$. 

The measurement is limited by the systematic uncertainty of IT-CsF2, including the density shift calculations, that is \num{2.3e-16} (Table~\ref{BudgetCs}). The uncertainties of IT-Yb1 (\num{1.9e-17}) and the microwave-to-optical conversion at the combs are negligible. With more than one year of data, the statistical uncertainty limited by the fountain instability is reduced below the systematic contribution.
In this way, we avoided the need to increase the fountain measurement time by using the H-maser as flywheel~\cite{Schwarz2020, Nemitz2021, Pizzocaro2020}, and measurements using the H-maser chain and the Optical Microwave chain are handled in the same way.

\subsection{Gravitational coupling of fundamental constants}
Clock comparisons have been used to constrain the variations of fundamental constants~\cite{Sanner2019,  Lange2021}.
For example, violations of the Einstein equivalence principle would result in changes in the frequency ratios of atomic clocks.
As a possible application, we investigate the coupling of our frequency measurement to the gravitational potential of the sun~\cite{Guena2012, Dzuba2017, Ashby2018, McGrew2019, Schwarz2020}.
We fit the data in Fig.~\ref{fig:data}a to $y = y_0 + A\cos\left(2\pi(t-t_0)/T_0\right)$, where $t$ is the date of measurement, $t_0$ is the date of the 2022 perihelion, and $T_0=\SI{365.26}{d}$ is the duration of the anomalistic year. We find $A = -0.0(2.0)\times 10^{-16}$. The amplitude of the annual variation of the gravitational potential is $\Delta\Phi = \num{1.65e-10} c^2$, where $c$ is the speed of light. The coupling coefficient of the Yb/Cs ratio to the gravitational potential is then found to be $\beta_\text{Yb,Cs} = A c^2/\Delta\Phi = -0.0(1.2)\times 10^{-6}$.
This result shows no violation of the Einstein equivalence principle and has a similar uncertainty while being largely independent to the result in Ref.~\cite{McGrew2019}. Following the same approach of Refs.~\cite{McGrew2019, Schwarz2020} it results in a constrain to the gravitational coupling of the electron-to-proton mass ratio of $k_{\mu} = -0.1(1.2)\times10^{-6}$. This result can be improved by the continuous long-term operation of frequency standards~\cite{Lange2021, Guena2012}.

\section{Conclusions}
\label{Conclusions}
\begin{figure}[h]
    \centering
    \includegraphics[scale=0.8]{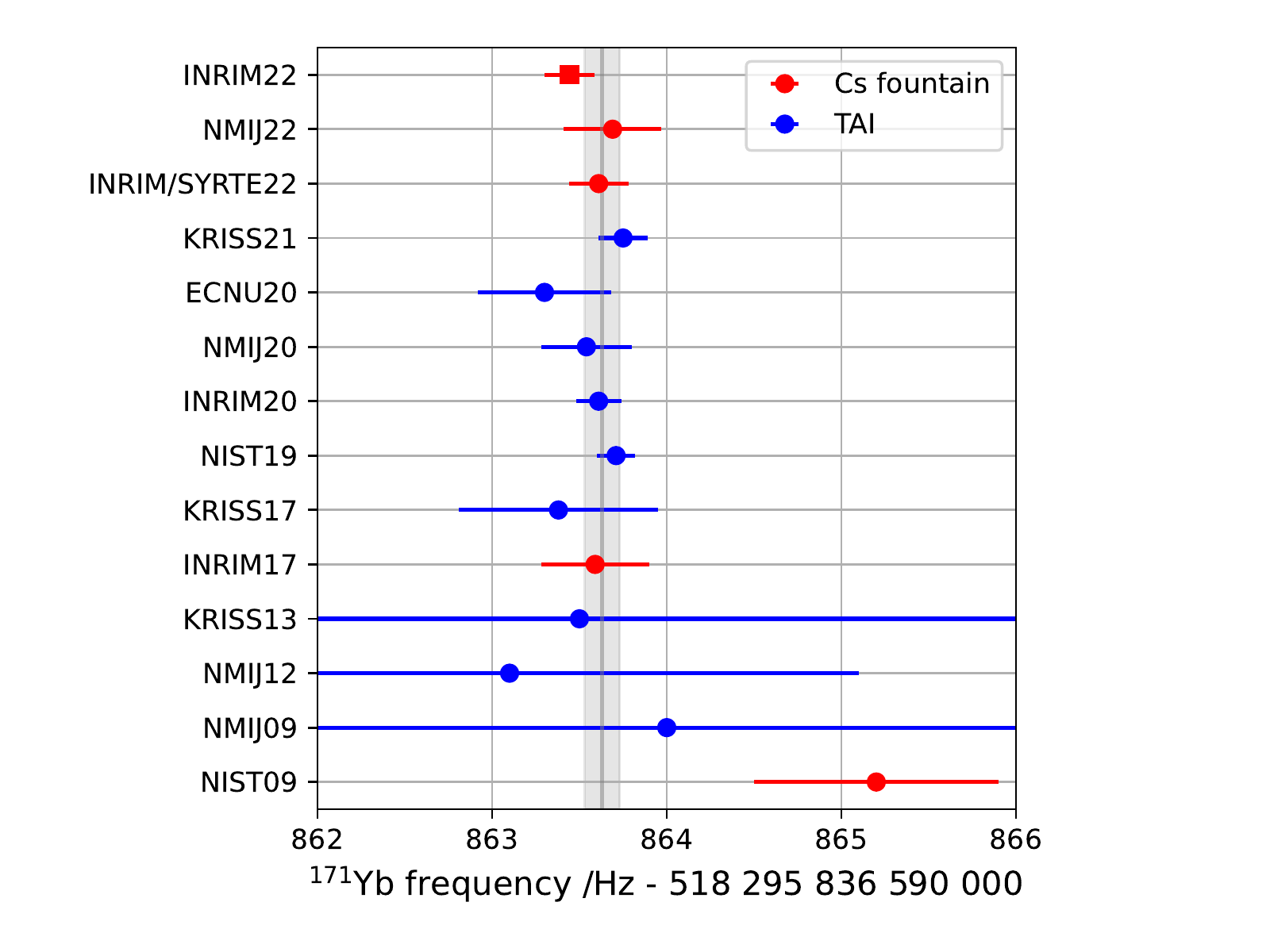}
    \caption{Absolute frequency of Yb performed in several metrological institutes worldwide from 2009 up to now. In red the measurements of Yb against a Cs fountain are shown, while in blue are reported the results of Yb frequency measurements obtained with a link to TAI. The result of the measurement described in this paper is shown as a red square. The grey solid line and shaded area are respectively the recommended secondary representation of the second and its uncertainty, given by CCTF.}
    \label{fig:Ybfreq}
\end{figure}

In this paper, we report the measurement of the clock transition of IT-Yb1 against the primary Cs fountain clock IT-CsF2. The measurement result is $f(\text{IT-Yb1}) = \SI{518295836590863.44(14)}{Hz}$ and the total fractional uncertainty, limited by the systematic uncertainty of IT-CsF2, is \num{2.7e-16}. This is the absolute frequency measurement of Yb against a Cs fountain with the lowest fractional uncertainty. We also show the updated uncertainty budget of IT-Yb1 that has a relative uncertainty of \num{1.9e-17}. In Fig.~\ref{fig:Ybfreq} our measurement reported as a red square is compared to the recommended secondary representations of the second and previous absolute frequency measurements~\cite{Lemke2009, Kohno2009, Yasuda2012, Park2013, Pizzocaro2017, Kim2017, McGrew2018, Pizzocaro2020, Kobayashi2020, Luo2020, Kim2021, Clivati2022a, Kobayashi2022}, showing a good agreement.
Moreover, we note that this measurement supersedes the local frequency measurement we reported in~\cite{Clivati2022a}, as it extends the analysed data period.

These results are a strong demonstration of the improvements achieved in the field of optical clocks and the consequent possibility of introducing a new definition of the second.

\ack
This work is supported by: the European Metrology Program for Innovation and Research (EMPIR) Projects 18SIB05 ROCIT and 20FUN08 Nextlasers, which have received funding from the EMPIR programme co-financed by the Participating States and from the European Union’s Horizon 2020 research and innovation programme.

\printbibliography

\end{document}